\documentclass[aps,article,nofootinbib,groupedaddress,twocolumn]{revtex4}
\usepackage{graphicx}
\usepackage{amsmath,amssymb}
\usepackage{hyperref}
\usepackage[usenames]{color}
\usepackage{url}

\hypersetup{
    colorlinks=true,
    linkcolor=black,
    citecolor=blue,
    urlcolor=black
}

\def\Journal#1#2#3#4{{#1} {\bf #2}, #3 (#4)}
\def\be{\begin{equation}}
\def\ee{\end{equation}}
\def\ba{\begin{eqnarray}}
\def\ea{\end{eqnarray}}

\def\f{\frac}
\def\l{\left}
\def\r{\right}

\newcommand\PRL{Phys.~Rev.~Lett.}
\newcommand\PRD{Phys.~Rev.~D}

\begin{document}

\title{How to optimally parametrize deviations from General Relativity \\
in the evolution of cosmological perturbations}
\date{\today}

\author{Levon Pogosian$^{1}$, Alessandra Silvestri$^{2}$, Kazuya Koyama$^3$, Gong-Bo Zhao$^{3}$}

\affiliation{$^1$Department of Physics, Simon Fraser University, Burnaby, BC, V5A 1S6, Canada \\
$^2$ Kavli Institute for Astrophysics and Space Research, MIT, Cambridge, MA 02139, USA  \\
$^3$Institute of Cosmology \& Gravitation, University of Portsmouth, Portsmouth, PO1 3FX, UK}

\begin{abstract}

The next generation of weak lensing surveys will trace the growth of large scale perturbations through a sequence of epochs, offering an opportunity to test General Relativity (GR) on cosmological scales.
We review in detail the parametrization used in MGCAMB to describe the modified growth expected in alternative theories of gravity and generalized dark energy models. We highlight its advantages and examine several theoretical aspects. In particular, we show that the same set of equations can be consistently used on super-horizon and sub-horizon linear scales. We also emphasize the sensitivity of data to scale-dependent features in the growth pattern, and propose using Principal Component Analysis to converge on a practical set of parameters which is most likely to detect departures from GR. The connection with other parametrizations is also discussed.
\end{abstract}

\maketitle

\section{Introduction}

Future weak lensing surveys, like the Dark Energy Survey~(DES)~\cite{DES}, Large Synoptic Survey Telescope (LSST)~\cite{LSST} and Euclid~\cite{Euclid} will measure lensing shear and galaxy counts at a sequence of redshifts, effectively mapping the evolution of the matter and metric perturbations. Much like the table top and solar system tests of General Relativity \cite{Will}, this will offer an opportunity to verify Einstein's equations that specify the way in which the matter, the gravitational potential and the space curvature are related to each other, and thus test the validity of GR on cosmological scales~\cite{latest-review}.

Several parametrizations of modified growth have been proposed in the literature~\cite{Wang:1998gt,Linder:2007hg,Caldwell:2007cw,Zhang:2007nk,Amendola:2007rr,Hu:2007pj,BZ08,Acquaviva:2008qp,ZPSZ08} and they can be separated into two types. The first type~\cite{Wang:1998gt,Linder:2007hg,Zhang:2007nk,Acquaviva:2008qp}, which can be called ``trigger'' parameters, are directly derived from observations with no need to evolve growth equations of motion. They are designed to detect a breakdown of the standard model, but their values do not necessarily have a physical meaning in any theory. The second type~\cite{Caldwell:2007cw,Amendola:2007rr,Hu:2007pj,BZ08,ZPSZ08} can be called ``model'' parameters. They have physical meanings and unique values in specific modified gravity theories and can be used to define a consistent set of equations with which to compare theoretical predictions to observations. Theoretical predictions for these
parameters in modified gravity and general dark energy models are studied in~\cite{Song:2010rm}.

Eventually, it will be possible to simultaneously fit a given parametrization to a combination of all data that probe the growth: CMB, weak lensing, galaxy counts, and peculiar velocities. One then needs a system of equations that is meaningful across a wide range of scales and redshifts. Since trigger parameters are constructed out of specific types of observables, they can lead to unnecessary complications and inconsistencies if used as model parameters for calculating predictions for other types of data. Also, the evolution of perturbations on super-horizon scales is governed by a set of consistency conditions which are separate from the sub-horizon dynamics. Namely, as shown in~\cite{Wands:2000dp,Bertschinger:2006aw}, in the absence of entropy perturbations, the space curvature defined on hyper-surfaces on uniform matter density, $\zeta$, must be conserved on scales outside the horizon in order to be consistent with the overall expansion of the universe. Hence, a consistent system of equations should decouple the super- and sub-horizon regimes. This separation of scales is made explicit in the Parameterized Post-Friedmann (PPF) Framework of~\cite{Hu:2007pj}, where a different systems of equations are used on super-horizon and sub-horizion scales. The advantage of the method advocated in this paper and used in MGCAMB~\cite{ZPSZ08,mgcamb} is that it employs a single system of equations across all linear scales, without sacrificing any of the important consistency conditions. The super-horizon and sub-horizon evolution decouple naturally, without having to be explicitly separated.

In principle, any modification of gravity can be formally described in terms of an effective dark energy fluid. We show that, by construction, our system of equations automatically conserves the energy-momentum of this fluid. Furthermore, the same condition that insures conservation of $\zeta$ outside the horizon implies adiabaticity of the effective fluid perturbations.

Aside from the consistency of the modified growth parametrization, an equally desired property is its simplicity. One wants to strike a balance between working with the fewest number of parameters possible, yet still having enough flexibility in the model to capture most of the significant information contained in the data. We show how one can determine the optimal number of parameters based on a Principal Component Analysis (PCA), such as one performed in~\cite{Zhao:2009fn}.

The rest of the paper is organized as follows. In Sec. \ref{sec:equations} we present a system of equations to
evolve linear perturbations in a general theory of gravity, demonstrate their consistency across all linear scales, and discuss their interpretation in terms of an effective dark energy. Sec. \ref{sec:pca} discusses a strategy for finding an optimal
number of parameters. In Sec. \ref{sec:others} we discuss the connection of our method to other parametrizations in the literature. We summarize in Sec. \ref{sec:summary}.

\section{Parameterized evolution of linear perturbations in a general theory of gravity}
\label{sec:equations}

In order to test gravity against the growth of cosmological perturbations in a model-independent way, one needs a set of equations to evolve linear perturbations without assuming GR.
The concept of the universe being described by a Friedmann-Robertson-Walker (FRW) background metric with small perturbations on large scales, and with the matter content that is covariantly conserved, is more general than GR and we adopt it as our starting point. We consider scalar metric perturbations about a FRW background in conformal Newtonian gauge, for which the line element reads
\be\label{metric}
ds^2=-a^2(\tau)\l[\l(1+2\Psi\r)d\tau^2-\l(1-2\Phi\r) d\vec{x}^2\r] \ ,
\ee
where $\Phi$ and $\Psi$ are functions of time and space. We will work in Fourier space and, for simplicity, only consider cold dark matter (CDM) in our equations. Our discussion can be generalized to include baryonic and radiation effects, which would be important at sufficiently early times. We assume adiabatic initial conditions and covariant conservation of the energy-momentum tensor of matter. At linear order, the conservation equations in Newtonian gauge are
\begin{eqnarray}
\label{matter-conservation}
\delta'+{k\over aH}v-3\Phi'&=&0, \\
v'+v-{k\over aH}\Psi&=&0 \ ,
\label{matter-continuity}
\end{eqnarray}
where $\delta$ is the energy density contrast, $v$ the irrotational component of the peculiar velocity, and primes indicate derivatives w.r.t.  $\ln a$. In what follows we will work with the gauge-invariant comoving density contrast
\begin{equation}
\Delta \equiv \delta + 3\f{aH}{k} v\,,
\label{Def:Delta}
\end{equation}
which is particularly convenient on super-horizon scales to avoid gauge artifacts.

In order to solve for the evolution of the four scalar perturbations $\{\Delta,v, \Phi,\Psi\}$ we need two additional equations, normally provided by a theory of gravity (such as GR) which specifies how the metric perturbations relate to each other, and how they are sourced by perturbations in the energy-momentum tensor. One can parametrize these relations as
\begin{eqnarray}
\label{gamma}
\frac{\Phi}{\Psi}&=&\eta(a,k), \\
\label{parametrization-Poisson}
k^2\Psi&=&-\f{a^2}{2M_P^2}\mu(a,k) \rho\Delta \ ,
\end{eqnarray}
where $M_P^2\equiv1/8\pi G$, and $\eta(a,k)$\footnote{Note that we are now using $\eta$ instead of $\gamma$ for the ratio of the potentials in order to avoid confusion with the growth index parameter $\gamma$~\cite{Wang:1998gt,Linder:2007hg}, and to be consistent with other literature~\cite{Amendola:2007rr,Bean:2009wj,Song:2010rm}.} and $\mu(a,k)$ are generic functions of time and scale; they will assume an explicit form once a theory is specified. For instance, equations for the growth of perturbations have been derived for Chameleon type scalar-tensor models~\cite{Khoury:2003aq} (such as $f(R)$~\cite{Capozziello:2003tk,Carroll:2003wy} theories) in \cite{Zhang:2005vt,Brax:2005ew,Song:2006ej,Bean:2006up,Pogosian:2007sw,Tsujikawa:2007xu}, and for the Dvali-Gabadadze-Porrati model (DGP)~\cite{Dvali:2000hr} and its higher-dimensional extensions~\cite{Dvali:2007kt} in~\cite{Koyama:2005kd,Song:2006jk,Song:2007wd,Cardoso:2007xc,Afshordi:2008rd}. Given the linear perturbation equations for a specific model, it is straightforward to the determine functions $\eta$ and $\mu$~\cite{BZ08,ZPSZ08,Afshordi:2008rd}. In principle, one could also use the functions $\mu$ and $\eta$ to represent dark energy perturbations~\cite{Bean:2001ys,Amendola:2003wa,Bean:2007ny,Koivisto:2005mm} or cosmological effects of massive neutrinos~\cite{Pastor}.

It should be noted that in a general theory of gravity
the superposition principle may not hold. Hence the dynamics of
large scale density fluctuations that one derives by perturbing  the
background solution at linear order need not be the same as the result of averaging
over small scale fluctuations. $N$-body simulations in scalar-tensor theories of
chameleon type~\cite{Oyaizu:2008tb, Oyaizu:2008tb, Li:2009sy, Zhao:2009ke}, such as f(R), and in higher-dimensional extensions of DGP~\cite{Schmidt:2009sg, Schmidt:2009sv, Khoury:2009tk} demonstrated that in those models on large scales one recovers the predictions of linear perturbation theory. However, this need not hold in all modified gravity model.
With this in mind, one can view $\mu$ as a purely phenomenological function which relates $\Delta$ and $\Psi$ and which could be non-linear in a particular modified gravity theory. In other words, in a non-linear theory $\mu$ may itself be a function of $\Delta$.

Together, Eqs. (\ref{matter-conservation})-(\ref{parametrization-Poisson}) provide a complete system for the variables $\{\Delta,v, \Phi,\Psi\}$.  They can be combined into a closed system for the variables $\Delta$ and $v$. It is instructive to write this system of equations in terms of dimensionless quantities; specifically, we will use $\rho=3M_P^2H_0^2\Omega_M a^{-3}$ and introduce the new variables:
\begin{eqnarray}
\label{dimensionless_var}
p \equiv {k \over aH} \ \ \ \ u\equiv {pv} \ \ \ \ E_m={\Omega_M \over a^3} \ \ \ \ E={H^2 \over H_0^2} \ .
\end{eqnarray}
Now, combining equations (\ref{matter-conservation})-(\ref{parametrization-Poisson}) and substituting the dimensionless variables (\ref{dimensionless_var}) we obtain
 the following system of equations:

\begin{eqnarray}
\label{Delta-pr-dimensionless}
\Delta'&=&{-{9 E_m \over 2E} \eta \mu \left[{1-\eta \over \eta} + {(\eta \mu)' \over \eta \mu}\right] \Delta
+\left[ 3{H' \over H} - p^2\right] u
\over p^2 +{9 E_m \over 2E} \eta \mu}\\
u'&=&-\left[2+{H' \over H} \right]u-{3 \over 2}{E_m \over E} \mu \Delta\,.
\label{v-pr-dimensionless}
\end{eqnarray}
Given the functions  $\mu$ and $\eta$, Eqs. (\ref{Delta-pr-dimensionless}) and (\ref{v-pr-dimensionless}) can be integrated numerically to find $\Delta$ and $u$. The potentials $\Phi$ and $\Psi$ can then be determined from (\ref{gamma}) and (\ref{parametrization-Poisson}).

Eqs. (\ref{Delta-pr-dimensionless}) and (\ref{v-pr-dimensionless}) hold for any $k$, and can be used to evolve cosmological perturbations from super-horizon to sub-horizon linear scales if $\mu(a,k)$ and $\eta(a,k)$ are provided. As we will show, these equations automatically ensure that the evolution of perturbations outside the horizon is independent of $\mu$ up to ${\cal O}(p^2/\mu \nu)$ terms, as required by the consistency with the FRW background (\ref{consistency}).

\subsection{The $\Lambda$CDM limit}

In the $\Lambda$CDM model, $\mu=1=\eta$, and Eqs. (\ref{Delta-pr-dimensionless}) and (\ref{v-pr-dimensionless}) simplify to
\begin{eqnarray}
\label{delta-gr}
\Delta'&=&{ 3{H' \over H} - p^2
\over p^2 +{9 E_m \over 2E} } u \\
u'&=&-\left[2+{H' \over H} \right]u-{3 \over 2}{E_m \over E} \Delta\,.
\label{v-gr}
\end{eqnarray}
At epochs when the radiation component can be ignored, $3(H'/H)=-9E_m/(2E)$ and Eq. (\ref{delta-gr}) becomes
\begin{eqnarray}
\label{delta-lcdm}
\Delta'=- u \ .
\end{eqnarray}
Combined with (\ref{v-gr}), this gives the usual second order equation
\begin{eqnarray}
\Delta''+\left[2+{H' \over H} \right]\Delta'-{3 \over 2}{E_m \over E} \Delta=0
\label{delta-double-gr}
\end{eqnarray}
which is scale-independent. In a modified gravity, however, the time evolution of $\Delta$ will be scale-dependent for a general $\mu$ and $\eta$. In particular, it will be scale-dependent even if $\mu$ and $\eta$ depend only on time, as long as $\mu' \ne 0$, as can be noticed from (\ref{Delta-pr-dimensionless}).

\subsection{Super-horizon evolution}

The consistency of the long wavelength perturbations with the FRW background requires that
\begin{equation}
\label{eq:zeta}
\zeta' \equiv (\delta-3\Phi)' = {\cal O}(p^2) \ ,
\end{equation}
on super-horizon scales (${p=k/(aH) \ll 1}$) for \emph{adiabatic} perturbations~\cite{Wands:2000dp,Bertschinger:2006aw}, where $\zeta$ is the curvature perturbation on hypersurfaces of uniform density~\cite{Bardeen:1980kt}. As pointed out in~\cite{Wands:2000dp},  Eq. (\ref{eq:zeta}) follows directly from matter conservation (Eq. (\ref{matter-conservation})), as long as the $kv/(aH)$ term is $O(p^2)$ (which is the case in GR and in all viable gravity models considered in the literature). The super-horizon conservation of $\zeta$ implies a second order differential equation for the metric potentials~\cite{Mukhanov:1990me,Bertschinger:2006aw}:
\be
\label{consistency}
\Phi''+\Psi'-\f{H''}{H'}\Phi'+\l(\f{H'}{H}-\f{H''}{H'}\r)\Psi={\cal O}(p^2)\ .
\ee
Therefore, once a relation between the two potentials is specified, i.e. once $\eta$ is given, Eq. (\ref{consistency}) is sufficient to solve for the evolution of super-horizon scale metric perturbations. Then Eq. (\ref{eq:zeta}) and the super-horizon limit of Eq. (\ref{matter-continuity}) can be used to infer $\delta$ and $v$.

In a multi-fluid system, the curvature perturbation on the uniform-total-density hypersurface, $\zeta$, remains constant on super-horizon scales if the uniform-density hypersurfaces for the different fluids coincide on superhorizon scales, i.e. if we have adiabatic initial conditions for the multi-fluid system. Since we are mostly interested in structure formation during matter domination and later,  we take $\zeta=\zeta_m$, and analyze the SH behavior of $\zeta$ in our framework (\ref{matter-conservation})-(\ref{parametrization-Poisson}). It is straightforward however to generalize the following discussion to the multi-fluid case.

It is easy to see that Eq. (\ref{consistency}) follows from the set of equations (\ref{Delta-pr-dimensionless}) and (\ref{v-pr-dimensionless}) as long $p^2/(\mu \eta)\rightarrow0$ in the small $p$ limit. Ideed, for $p \ll 1$, Eqs. (\ref{Delta-pr-dimensionless}) and (\ref{v-pr-dimensionless}) become
\begin{eqnarray}
\label{SH-conservation}
\Delta'&=&-\Delta \left[{1-\eta \over \eta} + {(\eta \mu)' \over \eta \mu}\right]
+{2 \over 3}{H'H \over \eta \mu E_m}u \\
u'&=&-\left[2+{H' \over H} \right]u-{3 \over 2}{E_m \over E} \mu \Delta\ .
\label{SH-continuity}
\end{eqnarray}
Combining them into a second order equation for $\Delta$, and using Eq. (\ref{parametrization-Poisson}), we obtain
\begin{multline}\label{consistency_Psi}
\Psi''+\l(2\f{\eta'}{\eta}-\f{H''}{H'}+\f{1}{\eta}\r)\Psi'+\l[\f{\eta''}{\eta}-\f{H''}{H'}\f{\eta'}{\eta}+\r.\\
\l.\l(\f{H'}{H}-\f{H''}{H'}\r)\f{1}{\eta}\r]\Psi=
{\cal O}\left(p^2 \over \mu \eta \right) \ . \end{multline}
This equation is equivalent to Eq. (\ref{consistency}), after Eq. (\ref{gamma}) is used to express $\Phi$ in terms of $\Psi$, as long as $\mu \eta$ does not approach $0$ faster than $p^2$. This requirement is likely to be satisfied in any reasonable model, given that $\mu=\eta=1$ in GR, and radical deviations from GR are typically discouraged by the data. Eqs. (\ref{matter-conservation})-(\ref{parametrization-Poisson}) are implemented in MGCAMB in synchronous gauge, and their super-horizon consistency was demonstrated in~\cite{ZPSZ08}.

Eq.~(\ref{consistency_Psi}) shows that in our framework the super-horizon
evolution of $\Phi$ and $\Psi$ is independent of $\mu$. Eq.~(\ref{eq:zeta}) implies that this is also the case for the
evolution of $\delta$ in the Newtonian gauge. On the other hand, $\mu$ and $\mu'$ appear explicitly in Eqs. (\ref{SH-conservation}) and (\ref{SH-continuity}), and it may seem
that $\mu$ could affect the evolution of super-horizon sized
perturbations. For instance, Eq. (\ref{SH-conservation}) implies that if one sets $\eta=1$ and $\mu'>0$, the growth of $\Delta$ will be suppressed. One way to see that this does not amount to an inconsistency with Eqs. (\ref{consistency}) and (\ref{consistency_Psi}) is to observe that $\Delta$
itself is ${\cal O}(p^2)$ on super-horizon scales\footnote{It is well-known that in synchronous gauge $\delta$ is ${\cal O}(p^2)$ on large scales~\cite{Ma:1995ey}. There the remaining gauge freedom is used to
set $v=0$ for CDM, so the synchronous gauge $\delta$ is the same as $\Delta$.}. Hence, whatever impact $\mu$
has on the super-horizon evolution of $\Delta$ can only be of order $O(p^2)$ and, therefore, is unobservable -- it is completely hidden by cosmic variance. We demonstrate this with an explicit example in a separate subsection.

\subsection{Sub-horizon evolution}

For $p \gg 1$ equations (\ref{Delta-pr-dimensionless}) and (\ref{v-pr-dimensionless}) become
\begin{eqnarray}
\Delta'&=&-u  \\
u'&=&-\left[2+{H' \over H} \right]u-{3 \over 2}{E_m \over E} \mu \Delta \ .
\end{eqnarray}
They can be combined into a second order equation:
\begin{eqnarray}
\Delta''+\left[2+{H' \over H} \right]\Delta'-{3 \over 2}{E_m \over E} \mu \Delta=0 \ ,
\label{delta-double-pr}
\end{eqnarray}
which has the same form as (\ref{delta-double-gr}) except for the rescaling of the Newton's constant by $\mu$. Therefore, the growth of $\Delta$ on sub-horizon scales is directly affected by $\mu(a,k)$, and is independent of $\eta$.

\subsection{An example}

\label{MGexample}

\begin{figure}[tbp]
\includegraphics[width=1.\columnwidth]{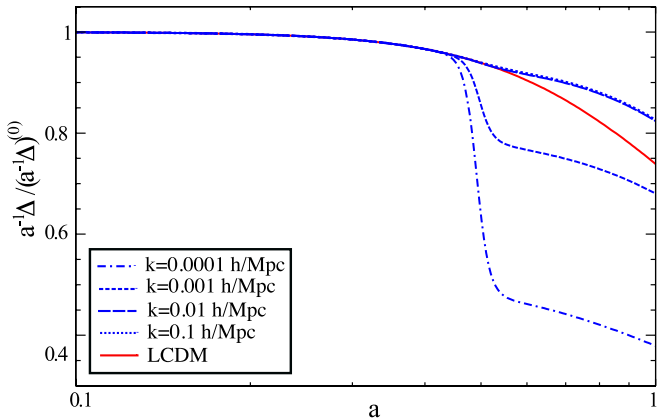}
\caption{The growth factor, $(\Delta(k,a)/a) /
(\Delta(k,a_i)/ a_i)$, for the  wavenumbers: $k=0.1$ (dotted line), $0.01$ (long-dashed),
$10^{-3}$ (short-dashed) and $10^{-4}$ (dot-dashed) h/Mpc as a function
of $a$ for $\mu(z)$ given by (\ref{mu-tanh}). The red
solid line is the $\Lambda$CDM solution, identical for each $k$. Note the approximately scale-independent
enhancement for sub-horizon modes ($k=0.1$ and $0.01$ h/Mpc) at
$z<1$ due to a rescaling of Newton's constant by $\mu$. The long
wavelength modes ($k=10^{-3}$ and $k=10^{-4}$ h/Mpc) experience a
suppression as expected from (\ref{SH-conservation}) for $\mu'>0$. However,
because $\Delta$ is ${\cal O}(p^2)$ for small
$k$, this suppression is concealed by cosmic variance as illustrated in Fig. \ref{fig:Pk}.}
\label{fig:growth}
\end{figure}

Let us consider a toy example to illustrate the theoretical arguments made in the previous subsections.
We choose $\eta=1$ and a scale independent form of
$\mu$:
\be
\label{mu-tanh}
\mu(z)=\left(1+\mu_0 \over 2
\right)+\left(1-\mu_0 \over 2 \right)\tanh\left[z-z_s \over \delta z
\right]  \ ,
\ee
which describes a transition from $\mu=1$ at $z>z_s$ to
$\mu=\mu_0$ at $z<z_s$, with a width set by $\delta z$. We set
$\mu_0=2$, $z_s=1$ and $\delta z=0.1$. Note that the transition in $\mu$ occurs on all scales, including those outside the horizon. If we interpret $\mu$ as a rescaling of Newton's constant, such a super-horizon variation is technically inconsistent with the Friedmann equation which sets the relation between the background metric and the background density. We choose this unphysical example on purpose, to demonstrate that, because Eqs. (\ref{matter-conservation})-(\ref{parametrization-Poisson}) are by design consistent with the background expansion on SH scales, the super-horizon variation in $\mu$ is undetectable (as expected from the arguments in the previous subsections). On sub-horizon scales, however, increasing the value of $\mu$ should result in enhanced clustering.

\begin{figure}[tbp]
\includegraphics[width=1.\columnwidth]{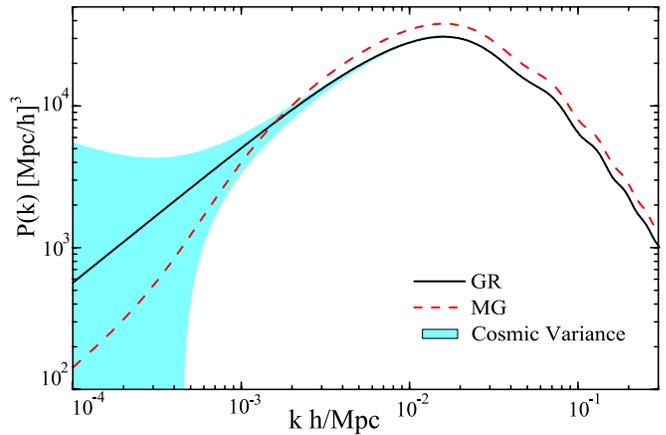}
\caption{The matter power spectrum at $z=0$ for $\Lambda$CDM (black solid curve) along with the associated cosmic variance (blue shaded region), and for the modified gravity (MG) example of Sec. \ref{MGexample} (red dashed curve). Note the potentially observable enhancement on sub-horizon scales, while the suppression due to the super-horizon variation of $\mu$ is completely hidden in cosmic variance.}
\label{fig:Pk}
\end{figure}

In Fig. \ref{fig:growth} we plot the gauge-invariant growth factor
\be
G(k,a)\equiv{\Delta(k,a)/a \over \Delta(k,a_i)/a_i} \ ,
\label{growth}
\ee
where $a_i$ is the scale-factor at some initial
time. We consider four values of $k$, ranging from $0.1$ to
$10^{-4}$ h/Mpc. The $\Lambda$CDM solution is scale-independent, as expected from Eqs. (\ref{v-gr}) and (\ref{delta-lcdm}), and is shown in the red solid line. However, the evolution becomes scale-dependent for
$\mu(z) \ne 1$, and the dashed and dotted blue lines show the solution for the four different scales in the case of $\mu$ given by (\ref{mu-tanh}).  One can notice an almost
scale-independent enhancement of growth for sub-horizon modes ($k=0.1$ and
$0.01$ h/Mpc) at $z<z_s$ as expected from Eq. (\ref{delta-double-pr}). On other hand, the modes that were fully
($k=10^{-4}$h/Mpc) or partially ($k=10^{-3}$h/Mpc) outside the horizon
when $\mu$ began growing, experience a suppression, as expected from (\ref{SH-conservation}) for $\mu'>0$. This suppression is not observable
because $\Delta$ is ${\cal O}(p^2)$ for small $p$ and is completely dominated by cosmic variance. This is demonstrated explicitly in Fig. \ref{fig:Pk}, where we plot the gauge-invariant matter power spectrum $P(k) \propto \Delta^2$, along with the associated cosmic variance.
\begin{figure}[tbp]
\includegraphics[width=1.\columnwidth]{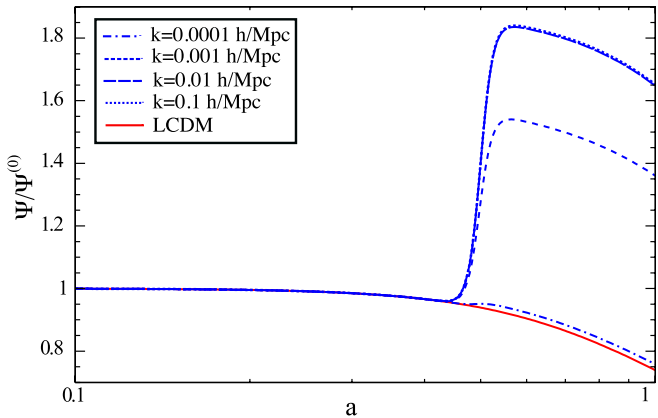}
\caption{The evolution of $\Psi(k,a)/\Psi(k,a_i)$ for
for the  wavenumbers: $k=0.1$ (dotted line), $0.01$ (long-dashed),
$10^{-3}$ (short-dashed) and $10^{-4}$ (dot-dashed) h/Mpc as a function
of $a$ for $\mu(z)$ given by (\ref{mu-tanh}), (as in Fig. \ref{fig:growth}). The red solid line is
the $\Lambda$CDM solution, which is scale-independent. Modifying
$\mu$ affects only the modes that cross the horizon. The super-horizon gravitational
potential, as expected, does not depend on the Poisson equation and, hence, the choice of $\mu$.}
\label{fig:psi}
\end{figure}

On super-horizon scales, the metric perturbations $\Psi$ and $\Phi$ determine observables such as the CMB temperature fluctuations.
Fig. \ref{fig:psi} shows the evolution of
$\Psi(k,a)/\Psi(k,a_i)$ for the model~(\ref{mu-tanh}) studied in this subsection. The red solid line is the scale-independent $\Lambda$CDM solution.
The effect of the transition in $\mu$ is profound on sub-horizon scales, but becomes smaller on larger
scales, and eventually disappears for super-horizon scales. The
independence of the SH evolution of the potentials on $\mu$
is physically expected; the Poisson
equation indeed should not play a role on SH scales, where the evolution of $\Psi$ and $\Phi$ is uniquely determined as a function of $\eta$~(\ref{consistency}).

\subsection{The effective dark fluid interpretation}
\label{sec:bianchi}

In an alternative gravity model, the Einstein equation can in general be written in terms of a functional of the metric, i.e. $E[g_{\mu \nu}] = 8 \pi G T_{\mu \nu}$. In the GR case $E[g_{\mu \nu}]$ corresponds to the Einstein tensor $G_{\mu\nu}$. One can always interpret the additional terms in $E[g_{\mu \nu}]$ as contributions from the energy-momentum tensor of an {\it effective dark fluid}. That is, formally we can always write
\be\label{eff_fluid}
E[g_{\mu \nu}]=G_{\mu \nu} - 8\pi G T^{\rm eff}_{\mu \nu}= 8\pi G T_{\mu \nu} \ .
\ee
In general, such a fluid will be imperfect, and its scalar perturbations will be characterized by a density contrast, velocity potential, pressure and a shear stress. Since the Bianchi identity is a geometrical property of the Riemann tensor, the contracted Bianchi identity
\be\label{Bid}
D^\mu G_{\mu\nu}= D^\mu \left(R_{\mu \nu}-{1\over 2} g_{\mu \nu}R \right)=0
\ee
holds independently of the form of the gravitational action at all orders in perturbation theory.
It is common to associate the contracted Bianchi identity with the diffeomorphism invariance of the action and with the conservation of energy-momentum of matter fields. Indeed, applying the contracted Bianchi identity to the Einstein equations in GR, one automatically obtains the covariant conservation of the energy-momentum tensor. Namely, $G_{\mu \nu} = 8 \pi G T_{\mu \nu}$ together with~(\ref{Bid}) implies $D^\mu T_{\mu\nu}=0$. Similarly, if we {\it assume} $D^\mu T_{\mu\nu}=0$\footnote{This assumption in part follows from our choice to work in the so-called Jordan frame, in which matter fields follow the geodesics of the metric. In other words, it amounts to the assumption that it is possible to write an action in which the matter Lagrangian is minimally coupled to the metric.}, as we did earlier, then the effective fluid will also be covariantly conserved at all orders in perturbation theory.

With two equations provided by the conservation of $T^{\rm eff}_{\mu \nu}$, at the linear level in scalar perturbations we have two remaining degrees of freedom required to specify the evolution of the dark fluid. These can be interpreted as the shear stress and the sound speed, which in our parametrized formalism of Eqs.~(\ref{matter-conservation})-(\ref{parametrization-Poisson}) can be determined once the functions $\mu(a,k)$ and $\eta(a,k)$ are specified. By construction, $T^{\rm eff}_{\mu \nu}$ will be conserved  for any choice of $\mu$ and $\eta$ and the evolution of perturbations in the effective fluid will be completely determined.

While the conservation of $T^{\rm eff}_{\mu \nu}$ does not impose any conditions on $\mu$ and $\eta$, we next show that there will be a constraint if we further require perturbations of the effective fluid to be \emph{adiabatic}. Using the version of~(\ref{eff_fluid}) linear in scalar perturbations,  the effective energy density and pressure perturbations can be written as
\be
\label{eff_density}
\rho_{\rm{eff}}\delta_{\rm{eff}}=-\rho\delta-\f{2M_P^2k^2}{a^2}\Phi-6M_P^2H^2\l(\Phi'+\Psi\r) \ ,
\ee
\begin{multline}
\label{eff_pressure}
\delta P_{\rm{eff}}=-\delta P+2M_P^2H^2\l[\Phi''+\l(\f{H'}{H}+3\r)\Phi'+\Psi'+\r.\\
\l.\l(2\f{H'}{H}+3\r)\Psi+\f{k^2}{3}(\Phi-\Psi)\r] \ ,
\end{multline}
where $\rho\delta$ and $\delta P$ are respectively the density and pressure perturbation of matter fields. Using Eqs.~(\ref{gamma}) and~(\ref{parametrization-Poisson}) we could further express these effective quantities in terms of the matter variables, $\eta$ and $\mu$, and their evolution would be fully specified for given $\eta$ and $\mu$.

Taking the super-horizon limit ($p\ll1$) we find
\begin{multline}\label{deltaPSH}
\delta P_{\rm{eff}}=c^2_{a(\rm{eff})}\delta\rho_{\rm{eff}}+2M_P^2H^2\l[\Phi''+\Psi'-\f{H''}{H'}\Phi'+\r.\\ \l.\l(\f{H'}{H}-\f{H''}{H'}\r)\Psi\r] \ ,
\end{multline}
where $c^2_{a(\rm{eff})}$ is the adiabatic speed of sound of the effective fluid, i.e.
\be\label{ad_speed}
c^2_{a(\rm{eff})}\equiv w_{\rm{eff}}-\frac{w'_{\rm{eff}}}{3(1+w_{\rm{eff}})}\,,
\ee
and where we have assumed that matter and the effective fluid have adiabatic modes on long wavelengths, i.e. that they share a uniform-density hypersurface.

From Eq.~(\ref{deltaPSH}) we see that in order for the effective fluid to have adiabatic pressure perturbations, i.e. $\delta P_{\rm{eff}}\rightarrow c^2_{a(\rm{eff})}\delta\rho_{\rm{eff}}$ on super-horizon scales, one must have
\be
\Phi''+\Psi'-\f{H''}{H'}\Phi'+\l(\f{H'}{H}-\f{H''}{H'}\r)\Psi={\cal O}(p^2) \ ,
\ee
which is equivalent to the consistency condition~(\ref{consistency}), derived from the conservation of the curvature perturbation $\zeta$ in absence of entropy perturbations.
In the previous subsection we have demonstrated that Eqs.~(\ref{matter-conservation})-(\ref{parametrization-Poisson}) satisfy the super-horizon conservation of $\zeta$ as long as $p^2/(\mu\eta)$ goes to zero in the $p\rightarrow0$ limt.

\section{How many parameters to fit?}
\label{sec:pca}

In the previous section we showed how two functions, $\mu$ and $\eta$, can be used in a consistent way to parametrize the linear growth of perturbations in a general modification of gravity. We have not, however, discussed the actual parametrization of the functions themselves. This is not an issue when testing a particular theory, where $\mu$ and $\eta$ have a specific time- and scale-dependence and can be specified with just a few parameters, e.g. via a direct use of Eqs.~(\ref{mu-chameleon}) or their equivalent~\cite{BZ08,ZPSZ08}. However, one may want to measure $\mu$ and $\eta$ from the data without necessarily assuming a particular class of models. The question then is how to strike a balance between simplicity, i.e working with as few parameters as possible, and allowing for enough flexibility in these functions to capture all of the information contained in the data~\footnote{The discussion of this section is specific to \emph{model} parameters and does not apply to triggers, which will be considered in Sec.\ref{sec:others}. Triggers are reconstructed directly from data, so there is no control on their form. However depending on their definition they might capture or not some important features in the data.}.

\begin{figure}[tbp]
\includegraphics[width=1.\columnwidth]{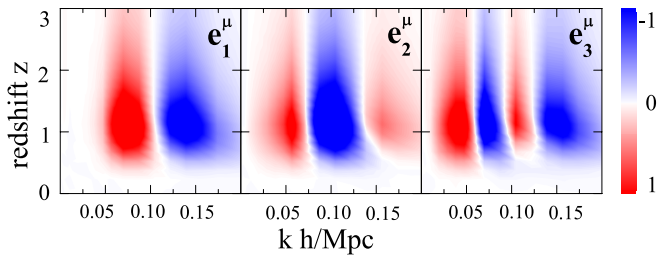}
\caption{The three best constrained eigenmodes of $\mu$ for DES,
adapted from Ref.~\cite{Zhao:2009fn}} \label{fig:mgpca}
\end{figure}

For example, one can discretize $\mu(k,z)$ and $\eta(k,z)$ on a grid in $(k,z)$ space and treat the values of the functions at each grid point, which we will call {\it pixel}, as independent parameters. In some of the earlier literature, $\mu$ and $\eta$ (or
a related set of parameters) were taken to be constants, or have two very wide pixels parameterizing a transition form GR at early times, to different values today~\cite{Daniel:2009kr,Bean:2009wj}. Scale-dependent variations have typically not been considered. This was motivated
primarily by simplicity. Namely, the idea is to start with the
simplest possible model and complicate it only if the fitted values
of this simple model parameters shows hints of departure from
$\Lambda$CDM. However, this logic may not always be appropriate
in modified gravity studies and might lead to missing on important information contained in the data.

Indeed, as was clearly shown in~\cite{Zhao:2009fn}, the growth data is much more sensitive to the {\it shape} of the functions $\mu(k,z)$
and $\eta(k,z)$, especially to their $k$-dependence, than to their
overall amplitude (or time-dependence).
In~\cite{Zhao:2009fn}, a Principal Component Analysis (PCA) was performed to find the
best constrained uncorrelated linear combinations of the $\mu(k,z)$
and $\eta(k,z)$ pixels -- the so-called {\it eigenmodes} -- for several surveys. For example, the
three  eigenmodes of $\mu$ best constrained by the Dark Energy Survey (DES)~\cite{DES} are shown in
Fig. \ref{fig:mgpca}.
Note that the best constrained mode has a node along the
$k$-direction, i.e. it is definitely not well approximated by a constant function of scale. More generally, in~\cite{Zhao:2009fn} the authors have found that the well-constrained modes of $\mu$ and $\eta$ for several surveys display a non trivial pattern in the $k$ direction, with no eigenmodes that have no nodes.

\begin{figure}[tbp]
\includegraphics[width=1.\columnwidth]{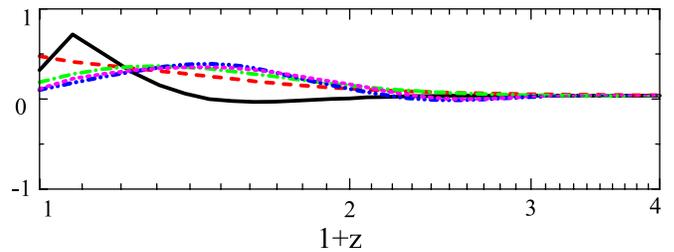}
\caption{The best measured eigenmodes of w from SNe (black solid), CMB (red
dash), galaxy counts (green dash dot), weak lensing(blue dash dot
dot) and combined (magenta short dash). This forecast result is
adapted from Ref.~\cite{PCA2}.}\label{fig:wpca}
\end{figure}

It is instructive to compare this with the findings of a PCA analysis on the dark energy equation of state $w(z)$~\cite{PCA1,PCA2}. In general, $w$ is a time-dependent function, and the simplest parametrization is that of a constant $w$. As it was found in~\cite{PCA1,PCA2}, almost every best constrained
mode of $w$ forecasted for different surveys has no nodes in $z$, as
shown in Fig. \ref{fig:wpca}. This means that observables are indeed most sensitive to an average value of $w$ and one can expect the tightest constraints on $w(z)$
when fitting a constant $w$, with constraints on variations in $w$
being weaker.

This is not the case for the functions $\mu(k,z)$ and
$\eta(k,z)$, especially in regard to their scale-dependence.
Because of
degeneracies with other parameters such as the $\Omega_{\rm m}$, $w$,
and $\sigma_8$, the growth observables are rather insensitive to the average
values of $\mu(k,z)$ and $\eta(k,z)$ over scale, and not too
sensitive to their time-dependence. On the contrary, data is rather sensitive
to their scale-dependence, as shown clearly by the modes of Fig. \ref{fig:mgpca}.
This is not surprising -- for example, redshift surveys primarily constrain the
shape of the power spectrum $P(k)$, with the amplitude being degenerate
with other parameters.

The above arguments prove that  when fitting $\mu(k,z)$ and
$\eta(k,z)$ to data, it is important for their parametrization to
be flexible enough to allow for some scale-dependence. Otherwise, we
would not be exploiting the true discovery potential of the data, and
might risk to miss on detecting departures from $\Lambda$CDM.

Still, while allowing for scale-dependence, one needs to choose an optimal pixelization, and the PCA method can be of help once again. Indeed, it allows to determine the optimal number of pixels
necessary to extract most of the information from the data. Having
too many pixels makes the fitting to data numerically difficult,
while having too few may lead to misleading results. A reasonable
way to arrive at the optimal number, would be to perform a Fisher Matrix forecast using the specifics of the particular
survey, such as DES and LSST, and then perform a PCA on the Fisher matrix. The best constrained eigenmodes can
then suggest the minimal number of pixels needed to describe them.
For example, the eigenmodes shown in Fig. \ref{fig:mgpca} suggest
that one should allow for the pixelization of $\mu$ to have at least two $k$-intervals, so that
it can describe a potential variation at $\sim$0.1 h$^{-1}$Mpc. They
also suggest that data cannot constrain sharp transition in time,
and one can have a single redshift pixel, with the transition at
$z\sim1.5$.

\section{Connection with other parametrizations}
\label{sec:others}

Several alternative parametrizations of modified linear growth have been proposed in the literature and they can be divided into two types. The first type consist in parameters that can be called {\it triggers}~\cite{Wang:1998gt,Linder:2007hg,Zhang:2007nk,Acquaviva:2008qp}. They can be derived directly from observations, with no need to evolve growth equation. Any disagreement between the observed trigger parameter and its value in the $\Lambda$CDM model would indicate some sort of a modification of growth. The second type consists of parameters and/or functions which can be called {\it model} parameters and are introduced to play a role similar to our functions $\mu$ and $\eta$. Namely, they are used to build a system of equations for the evolution of metric and density perturbations. Unlike the trigger types, which are directly calculated from the observables, the model parameters are measured by fitting the model that they define to data. Both types of parameters have their purpose. Working with functions like $\mu$ and $\eta$ gives one a consistent set of equations with which to compare theoretical predictions to observations. A measurement of a particular form of scale- or time-dependence of these functions could directly rule out, or point to a theory. The trigger parameters, on the other hand, are designed to detect a breakdown of the standard model, but their value does not necessarily have a physical meaning in any theory.

For example, the commonly used trigger parameter
$\gamma$~\cite{Wang:1998gt,Linder:2007hg} is defined via
\be
f \equiv {d \over d \ln a } \left( \ln {\Delta(k,a) \over \Delta(k,a_i)} \right) = \Omega_m(a)^{\gamma} \ ,
\label{def:gamma}
\ee
where $\Omega_m(a)=E_m/E$. As shown in~\cite{Wang:1998gt}, $\gamma=6/11\approx 0.55$ provides a solution to Eq. (\ref{delta-double-gr}) if one neglects terms ${\cal O}[(1-f)^2]$ (or ${\cal O}[1-\Omega_m(a)]^2$). Since most of the information on linear clustering is expected to come from epochs before the matter-$\Lambda$ equality, the approximation involved in this parametrization is not unreasonable.
In principle, $f$ can be extracted by observing clustering at several redshifts, while $\Omega_m(a)$ can be measured from background expansion probes, like CMB and supernovae. Then a significant deviation of the observed $\gamma$ from its predicted value of $0.55$ would indicate a breakdown of $\Lambda$CDM.

The trigger parameters, such as $\gamma$ can in principle be used also as model parameters. For example,  given $\gamma$, one can integrate Eq. (\ref{def:gamma}) to find $\Delta$. To find the gravitational potentials $\Phi$ and $\Psi$, their relation to each other and to $\Delta$ must be specified additionally. In other words, one needs to provide a total of three functions instead of two (e.g. $\mu$ and $\eta$ introduced earlier), or assume that one or more of the Einstein equations are valid.  By working with $\gamma$, one essentially no longer assumes conservation of matter energy-momentum. Alternatively, one can use Eqs. (\ref{delta-double-pr}) (since $\gamma$ is only meant to be used to characterize the growth on sub-horizon scales) and (\ref{def:gamma}) to express $\mu$ in terms of $\gamma$:
\be
\mu=\frac{2}{3}\Omega_m^{\gamma-1}\left[\Omega_m^{\gamma}+2+\frac{H'}{H}+\gamma\frac{\Omega_m'}{\Omega_m}+\gamma'\ln\left(\Omega_m\right)\right] \ .
\label{mu-gamma}
\ee
For a flat $\Lambda$CDM background eq.(\ref{mu-gamma}) simplifies to 
\be
\mu=\frac{2}{3}\Omega_m^{\gamma-1}\left[\Omega_m^{\gamma}+2-3\gamma+3\left(\gamma-\tfrac{1}{2}\right)\Omega_m+\gamma'\ln\left(\Omega_m\right)\right] \ .
\label{mu-gamma_LCDM}
\ee
Then, for a given $\gamma$, and with an additional specification of $\eta$, one can use (\ref{mu-gamma}) in Eqs. (\ref{matter-conservation})-(\ref{parametrization-Poisson}) to find consistent solutions for the linear perturbations. Note that with $\gamma=6/11$ substituted into (\ref{mu-gamma_LCDM}), $\mu$ is not a constant, but a slowly varying function evolving from $\mu=1$ during matter domination to $\mu \approx 1.04$ today (for standard $\Lambda$CDM parameters)
due to the ${\cal O}[(1-f)^2]$ corrections.

Given that trigger parameters are designed in terms of a specific observable and for a limited purpose, they can lead to unnecessary complications and inconsistencies if used as model parameters for calculating predictions for other types of data. When performing a global fit to all available data, one needs a consistent system of equations for calculating predictions for all of the observables: CMB, weak lensing, galaxy counts and peculiar velocities.

As discussed above, in addition to the conservation of energy-momentum one needs to specify two functions relating $\Delta$ to the potentials, and the two potentials to each other. The actual choice of the two relations that define such two functions is no unique. However, working with $\eta$ (\ref{gamma}) and $\mu$ (\ref{parametrization-Poisson}) has certain advantages. One is that on super-horizon scales only $\eta$ affects the perturbations, while $\mu$ naturally becomes irrelevant. Also, on sub-horizon scales, only $\mu$ affects the growth of matter over-densities. The physics determining the evolution of perturbations on super-horizon scales is not necessarily related to their sub-horizon dynamics. This distinction is made explicit in the Parameterized Post-Friedmann (PPF) Framework of~\cite{Hu:2007pj}, where different systems of equations are used on super-horizon and sub-horizon scales. On super-horizon scales one must provide only $g=(1-\eta)/(1+\eta)$, while on sub-horizon scales one needs $g$ and $f_G \equiv-1-a^2\rho\Delta/(M_P^2k^2(\Phi+\Psi))$. A transition between the two regimes needs to be additionally specified in a way that is consistent with the conservation of energy-momentum. The advantage of the PPF is that it  does not need even weak assumptions on $g$ and $f_G$ in order to satisfy the consistency conditions, while our $\mu$ and $\eta$ technically must obey $p^2/(\mu \eta) \rightarrow 0$ for small $p$. The latter, however, is an extremely mild assumption, that is likely to be satisfied in any reasonable modified gravity theory. In turn, the benefit of our approach is that one evolves the same system of equations on all linear scales, with no need for an additional transition function.

Sometimes, it is convenient to define the function $\Sigma(k,a)$
\be\label{Sigma}
\Sigma(a,k)\equiv-\f{k^2M_P^2\l(\Phi+\Psi\r)}{\rho a^2\Delta} = \frac{(1+\eta)}{2} \mu,
\ee
which is directly related to the lensing potential $\Phi+\Psi$ (much like $f_G$ in PPF). As such, weak lensing (WL) measurements are sensitive to $\Sigma$ as well as the Integrated Sachs-Wolfe effect (ISW), which is determined by the time variation of the lensing potential. Thus $\Sigma$ is more directly constrained by these observations than $\eta$ or $\mu$. Another advantage of this function is that in the popular scalar-tensor/Chameleon theories, as well as in higher-dimensional theories, such as DGP~\cite{Dvali:2000hr}, this function assumes a simple expression and is effectively $1$ on small scales~\cite{Song:2010rm}.
Specifically, in Chameleon type theories in which the scalar field $\phi$ couples to CDM with a coupling $\alpha(\phi)$ one has\begin{eqnarray}
&&\mu(a,k)=\f{m^2a^{2}+\l(1+\f{1}{2}{\alpha_{\phi}}^2\r)k^2}{m^2a^{2}+k^2} e^{-\alpha(\phi)/M_P} \\
&&\eta(a,k)=\f{m^2a^{2}+\l(1-\f{1}{2}{\alpha_{\phi}}^2\r)k^2}
{m^2a^{2}+\l(1+\f{1}{2}{\alpha_{\phi}}^2\r)k^2}\,,
\label{mu-chameleon}
\end{eqnarray}
where $\alpha_{\phi}\equiv d\alpha/d\phi$ and $m$ is a time-dependent effective mass scale. Note that both $\mu$ and $\eta$ are functions of time and scale. On the other hand, $\Sigma$ is only a function of $time$:
\be
\Sigma(a,k)=e^{-\alpha(\phi)/M_P} \, .
\ee
In the DGP model and its higher dimensional generalizations we have~\cite{Afshordi:2008rd}
\be
\Sigma(a,k)= {1 \over 1+\left(a/kr_c \right)^{2(1-\alpha)}} \,
\ee
where $r_c$  is a characteristic lengthscale of the model, while $\alpha=1/2$ for DGP and $0$ for two or more extra-dimensions. Since on super-horizon scales one function is sufficient to fully describe the evolution of perturbations, it is best to use $\Sigma$ in pair with $\eta$, and not with $\mu$. This way, it will automatically becomes negligible on super-horizon scales and $\eta$ will be the only important function.

In Caldwell {\it et al.}~\cite{Caldwell:2007cw} the authors introduced  what they called the {\it gravitational slip}, i.e. a function $\varpi$ parametrizing the difference between the gravitational potentials as
\be\label{grav_slip}
\varpi\equiv\f{\Psi}{\Phi}-1=\f{1-\eta}{\eta}\, .
\ee
It should be used in combination with another function parametrizing the relation of metric potentials to matter density contrast, e.g. $\mu$. Also, generally one should allow for it to be scale-dependent.

Amendola {\it et al.}~\cite{Amendola:2007rr} use a function equivalent to $\varpi$ (\ref{grav_slip})  in combination with either $\Sigma$ (\ref{Sigma}) or $Q$, defined as
\be\label{Q}
Q\equiv-\f{k^2M_P^2 \Phi }{\rho a^2\Delta}=\mu\eta=\f{2\Sigma\eta}{1+\eta}\,.
\ee
As for $\Sigma$, an appropriate pair to describe modified growth would be $(Q,\eta)$.

Different observables will probe different combinations of the density contrast and metric potentials corresponding to one or more of the parametrizing functions.
For example clustering, i.e. the growth of $\Delta$, responds to the potential $\Psi$, and therefore to the function $\mu$. However, it is also affected by the magnification bias, i.e.  the lensing of the clustering pattern of the background sources by the intervening gravitational potentials. Hence, any realistic observation of clustering will be sensitive also to $\Sigma$ (and hence $\eta$) as well as $\mu$. Weak lensing and the ISW effect  probe the sum of the potentials ($\Phi+\Psi$) and, therefore, are directly related to the function $\Sigma$; of course they also constrain $\mu$ and $\eta$, with stringent constraints on $\mu$ as shown in~\cite{Zhao:2009fn}. Finally, peculiar velocities respond directly to the potential $\Psi$, therefore to the function $\mu$.

In summary, given that two functions are necessary to parametrize departures from LCDM in the growth pattern, one needs to decide which pair among the several functions described above, is best suited for a given analysis and for presenting the corresponding results. Our choice of the pair ($\eta$, $\mu$) has been guided by the fact that on super-horizon scales $\mu$ naturally becomes irrelevant and we are left with only one function, $\eta$, as expected on those scales. Notice also the importance of $\mu$ being defined through the Poisson equation involving $\Delta$ and not $\delta$; this allows for $\mu$ to be defined consistently on all scales and to go exactly to unity on all scales for LCDM.  From ($\eta$, $\mu$), we can derive other parameters which may be more suitable for interpreting observational constraints. For example, we can choose to describe departures from GR in terms of ($\Sigma$, $\mu$) because
WL measurements and the ISW effect are sensitive to $\Sigma$, and $\mu$ can be determined from peculiar velocity measurements, while $\eta$ is not probed directly by any observable and would be highly degenerate with $\mu$ or $\Sigma$.

\section{Summary}

As of today, the cosmological concordance model,  $\Lambda$CDM,  provides a good fit to all available observations. However, upcoming and future weak lensing surveys will bring a significant improvement in cosmological data sets and will offer an unprecedented opportunity to test GR on cosmological scales. As demonstrated in~\cite{Zhao:2009fn}, the currently available clustering information pales in comparison to what we will learn from a survey like LSST. A number of alternative gravity theories that are currently indistinguishable from $\Lambda$CDM will be tested, potentially providing clues about the physics causing cosmic acceleration. Even if no hints of new physics are observed, at the very least we can directly confirm the validity of GR at epochs and scales on which it has not been tested before.

This paper is a step towards building an optimal framework for testing GR. We showed how a single system of equations can be used consistently to evolve perturbations across all linear scales. These equations are implemented in MGCAMB~\cite{ZPSZ08,mgcamb}, a publicly available modification of CAMB~\cite{camb}, which can be used to evaluate CMB, weak lensing, number counts, and peculiar velocities spectra for a choice of functions $\mu$ and $\eta$. We have also argued, based on the scale- and redshift-dependent patterns of the best constrained eigenmodes, that future surveys will primarily probe the scale-dependence, and not so much the overall normalization or the time-dependence of these functions. Hence, in order to fully exploit the discovery potential of data, parametrized modifications of gravity must allow for scale-dependence.

While we have not addressed it in this paper, the galaxy counts only trace the underlying density field up to a bias factor. As explained in~\cite{Hui:2007zh}, any scale-dependence in the linear growth, will result in a scale-dependence of the linear bias. The latter however, is directly related to the scale-dependence of $\Delta(k,z)$ and therefore can be determined once $\mu(k,z)$ is specified.

Most of the clustering information comes, and will continue to come, from scales that have crossed into the non-linear regime, not described by the linear parametrization of Sec. \ref{sec:equations}. One could test gravity on non-linear scales by designing trigger parameters that would indicate a breakdown of GR. Alternatively, $N$-body simulations and higher order perturbative expansions~\cite{Koyama:2009me} can be used to constrain particular types of modified gravity theories \cite{Beynon:2009yd}. 

The precise accuracy of the tests will strongly depend on the ability to control the systematic effects, and their extent will not be fully known until experiments begin operating. Some preliminary estimates of the effect of systematics on cosmological test of GR have been reported in~\cite{Zhao:2009fn} and a more comprehensive analysis will be presented in~\cite{long_pca}.

{\it Note added: while we were preparing the manuscript for submission, a paper appeared on arXiv \cite{Daniel:2010ky} where some of the issues raised in this paper are also discussed.}

\label{sec:summary}

\acknowledgements We thank Ed Bertschinger, Robert Crittenden, Andrei Frolov, Luca Giomi, Lado Samushia, Yong-Seon Song, and David Wands for useful discussions. LP is supported by an NSERC Discovery grant and funds from SFU, AS by the grant NSF AST-0708501, KK is supported by the European Research Council (ERC), Research Councils UK and STFC. GZ is supported by the ERC.

\end{document}